\documentclass[aps,prm,reprint, superscriptaddress]{revtex4-2}
\usepackage{blindtext}
\usepackage{graphicx}
\usepackage{xcolor}
\usepackage{dcolumn}
\usepackage{bm}
\usepackage[utf8]{inputenc}

\begin{document}
	
	\preprint{APS/123-QED}
	
	\title{Electronic Coherence Evolution at the Nearly Commensurate–Incommensurate CDW Boundary of 1\textit{T}-TaS$_2$}
	
	\author{Turgut Yilmaz}
	\email{trgt2112@gmail.com}
	\affiliation{Department of Physics, Xiamen University Malaysia, Sepang 43900, Malaysia}
	\affiliation{Department of Physics, University of Connecticut, Storrs, CT 06269, USA}
	
	\author{Yi Sheng Ng}
	\affiliation{Department of New Energy Science and Engineering, Xiamen University Malaysia, Sepang 43900, Malaysia}
	
	\author{Menka Jain}
	\affiliation{Department of Physics, University of Connecticut, Storrs, CT 06269, USA}
	\affiliation{Institute of Materials Science, Storrs, University of Connecticut, 06269, USA}
	
	\author{Xiao Tong}
	\affiliation{Center for Functional Nanomaterials, Brookhaven National Lab, Upton, New York, 11973, USA}
	
	\author{Thipusa Wongpinij}
	\affiliation{Synchrotron Light Research Institute (Public Organization), Ministry of Higher Education Science Research and Innovation, Nakhon Ratchasima 30000, Thailand}
	
	\author{Pat Photongkam}
	\affiliation{Synchrotron Light Research Institute (Public Organization), Ministry of Higher Education Science Research and Innovation, Nakhon Ratchasima 30000, Thailand}

	\author{Anil Rajapitamahuni}
	\affiliation{National Synchrotron Light Source II, Brookhaven National Lab, Upton, New York 11973, USA}
	
	\author{Asish K. Kundu}
	\affiliation{National Synchrotron Light Source II, Brookhaven National Lab, Upton, New York 11973, USA}

	\author{Jin-Cheng Zheng}
	\affiliation{Department of Physics, Xiamen University, Xiamen 361005, P. R. China}
	\affiliation{Department of Physics, Xiamen University Malaysia, Sepang 43900, Malaysia}
	
	\author{Elio Vescovo}
	\affiliation{National Synchrotron Light Source II, Brookhaven National Lab, Upton, New York 11973, USA}

	\date{\today}

	\begin{abstract}
		Transition-metal dichalcogenides host a variety of charge-density-wave phases that  couple lattice, charge, and correlation effects. In 1\textit{T}-TaS$_2$, the commensurate and nearly commensurate states are well characterized, yet the transition near 350~K into the incommensurate phase has lacked direct momentum-resolved insight. Here we use temperature-dependent angle-resolved photoemission spectroscopy to track the electronic structure across this transition. We observe a suppression of quasiparticle spectral weight at the Brillouin-zone center, coincident with the transport anomaly, but without clear evidence of a full band gap opening.The transition appears to involve momentum-dependent redistribution of spectral weight, consistent with a loss of coherence that reshapes the Fermi surface while leaving conduction dispersions largely intact. These results suggest that the nearly commensurate–incommensurate transition may not align with a conventional metal–insulator transition picture, but rather as an electronic reconstruction driven by loss of coherence. Our work provides new microscopic insight into the resistivity anomaly near room temperature and may guide design principles for collective electronic switching in Transition-metal dichalcogenides.
	\end{abstract}

	\maketitle
	

	Transition-metal dichalcogenides (TMDCs) have emerged as exemplary systems for exploring correlated electronic phenomena such as charge-density-wave (CDW), Mott insulating behavior, and superconductivity~\cite{manzeli20172d,wilson1975charge,rossnagel2011origin,kogar2017signatures,thorne1996charge}. Their quasi-two-dimensional structure enables strong coupling between lattice and electronic degrees of freedom, producing rich phase diagrams with multiple temperature-dependent electronic and structural transitions. Among them, 1\textit{T}-TaS$_2$ stands out due to its complex sequence of CDW phases and the interplay with electron correlation effects. Upon cooling, it undergoes transitions from a high-temperature incommensurate (IC)-CDW phase to a nearly commensurate (NC)-CDW phase near 350 K, and finally to a commensurate (C)-CDW phase below ~180 K that is often associated with a Mott insulating state~\cite{fazekas1979electrical,sipos2008mott,perfetti2005unexpected,fazekas1980charge}.
	
	While the commensurate low-temperature phase has been extensively investigated, the microscopic nature of the NC-IC transition remains scarcely studied. Transport studies reveal a resistivity jump near 350~K, indicating a reorganization of the electronic states~\cite{fazekas1979electrical}. However, despite extensive ARPES investigations of 1\textit{T}-TaS$_2$, previous work has almost exclusively focused on the low-temperature commensurate phase~\cite{ang2012real,ritschel2015orbital,wang2020band,jung2022surface,yang2022visualization,wang2024dualistic}, while NC-IC boundary has been comparatively less explored in momentum space. As a result, the evolution of the $\Gamma$-centered band near the Fermi level (E$_F$) and its role in the transport anomaly have remained unresolved. Addressing this open question is scientifically significant and technologically relevant, since the sharp and reversible resistivity change near room temperature offers promising opportunities for device architectures based on collective electronic switching, including ultrafast electronic switches and functional materials for information processing~\cite{perfetti2006time,hellmann2012time,yoshida2015memristive,stojchevska2014ultrafast,ligges2018ultrafast}.
	
	In this work, we carry out a comprehensive temperature-dependent ARPES study of \textit{1T}-TaS$_2$ across the NC–IC transition during warm up, addressing a regime previously unexplored in momentum space. We observe a notable weakening of quasiparticle spectral weight near $E_F$ around 350~K, coinciding with the resistivity jump seen in transport. Importantly, this suppression does not correspond to the opening of a full bandgap, indicating that the transition is not a conventional metal–insulator transition. Instead, our results reveal a momentum-selective loss of coherence in the low-energy electronic states and a transformation of the Fermi surface topology. This electronic reconstruction, driven by the structural reconfiguration of the CDW, underscores the profound coupling between lattice and electronic degrees of freedom and provides new microscopic insight into CDW transitions.

	\begin{figure*}[htp]
		\centering
		\includegraphics[width=0.9\textwidth]{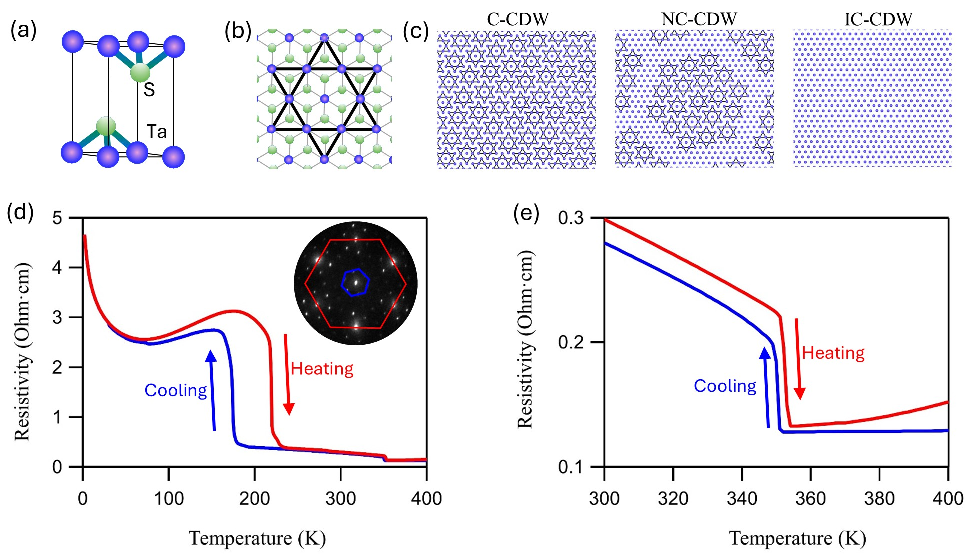}
		\caption{
			(a) Schematic of the undistorted trigonal (normal) phase. (b) Star-of-David distortion representative of the C-CDW phase. (c) Schematic phase diagram showing the evolution of CDW phases—C-CDW, NC-CDW, and IC-CDW—as a function of temperature. (d) Temperature-dependent resistivity measured during cooling and warming, displaying hysteresis between ~180 K and 240 K, characteristic of the first-order NC–C transition. Inset: LEED pattern obtained at 300~K with 50~eV electrons, showing sharp diffraction spots from the bulk crystal lattice together with additional satellite peaks arising from the commensurate CDW modulation. The red hexagon outlines the primary hexagonal diffraction from the bulk lattice, while the blue hexagon highlights the CDW-induced $\sqrt{13} \times \sqrt{13}$ superlattice associated with the Star-of-David reconstruction, demonstrating the well-ordered CDW state and high crystal quality. (e) Magnified view of the resistivity near 350 K, where the slope change supports the presence of a CDW-related structural or electronic transition.
		}
	\end{figure*}

	\section*{CDW Phase Sequence and Transport Response}
	
	Fig.~1 presents the crystal structure and transport behavior of 1\textit{T}-TaS$_2$. Fig.~1(a) and (b) depict schematic representations of the undistorted trigonal (normal) phase and the Star-of-David distortion characteristic of the C-CDW phase, respectively. In the latter, 13 Ta atoms within each cluster undergo periodic displacements, forming a $\sqrt{13} \times \sqrt{13}$ superlattice that strongly modulates the electronic potential~\cite{fazekas1979electrical,rossnagel2011origin}. The overall temperature evolution of the CDW order is summarized in Fig.~1(c), which traces the sequence of C-, NC-, and IC-CDW phases upon heating~\cite{wilson1975charge,sipos2008mott,ritschel2015orbital}. A representative low energy electron diffraction pattern (LEED) at 300 K in the NC-CDW phase (inset of Fig.~1(d)) shows clear CDW-related diffraction spots~\cite{von2019surface}, confirming the expected periodic modulation and the high crystalline quality of the samples.  
	
	The corresponding temperature-dependent resistivity (Fig.~1(d)) exhibits pronounced hysteresis between approximately 180~K and 240~K, reflecting the first-order C-NC transition~\cite{fazekas1979electrical}. At higher temperatures, a second anomaly appears near 350~K, marking the transition into the IC-CDW phase. The relatively sharp change in resistivity at this temperature points to a modified electronic structure near $E_F$. Fig.~1(e) magnifies the resistivity curve near 350~K, where a distinct slope change further supports the presence of a CDW-related transition. Beyond their fundamental importance, such macroscopic signatures have been shown to enable collective switching phenomena, including ultrafast nonthermal switching~\cite{stojchevska2014ultrafast} and memristive behavior in thin flakes~\cite{yoshida2015memristive}.

	\begin{figure*}[t]
		\centering
		\includegraphics[width=0.9\textwidth]{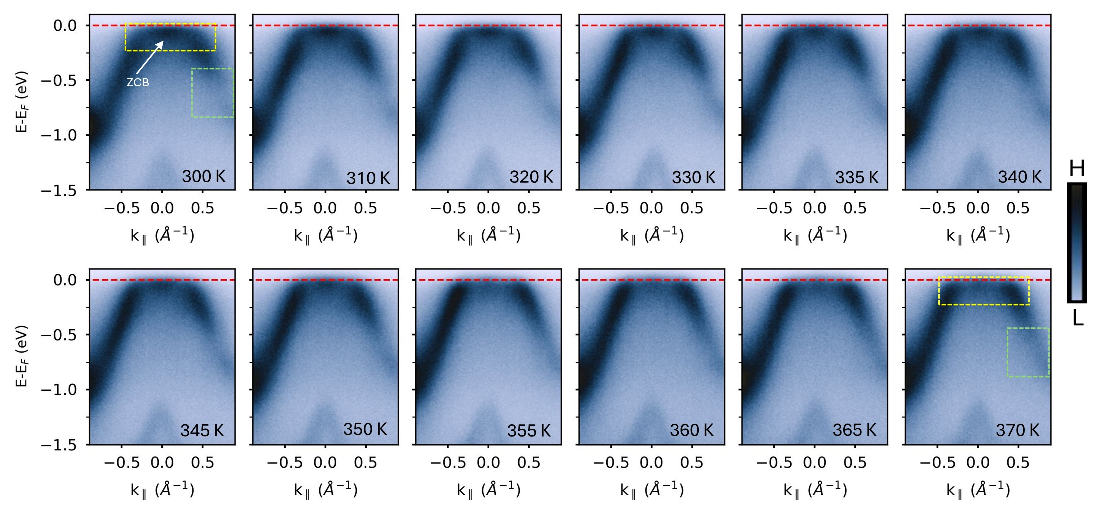}
		\caption{
			ARPES measurements along the high-symmetry $M$–$\Gamma$–$M$ direction reveal the evolution of low-energy electronic structure from 300 K to 370 K. The data were acquired using 92 eV photon energies. At 300 K, the spectrum shows two dispersive bands and a ZCB (highlighted by the yellow dashed square), with local energy gaps indicative of periodic lattice distortions (green dashed square). As the temperature crosses 350 K, these energy gaps nearly diminish, and ZCB loses spectral weight. The color bar represents the ARPES intensity with endpoints marking high (H) and low (L) spectral weigh.
		}
	\end{figure*}
	
	\begin{figure*}[t]
		\centering
		\includegraphics[width=0.9\textwidth]{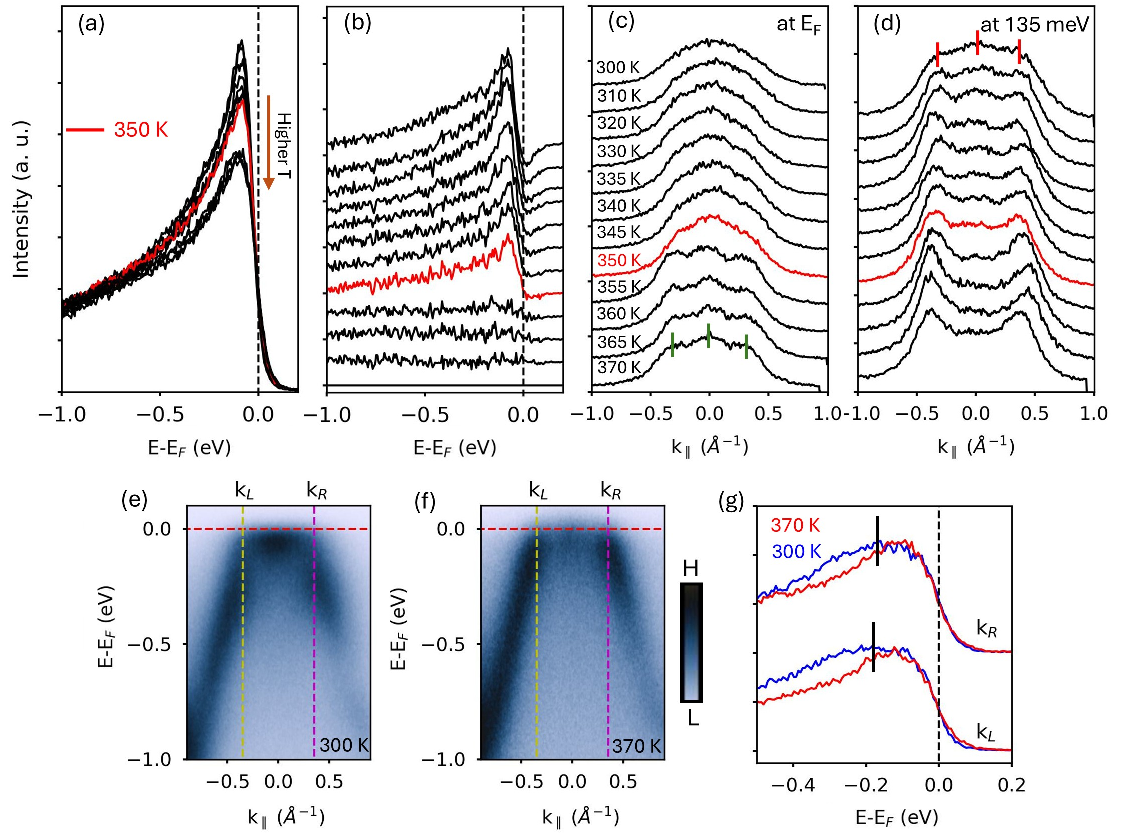}
		\caption{
			(a) Temperature dependent EDCs at the $\Gamma$ point. The quasiparticle peak associated with ZCB is suppressed above ~350 K. (b) Differential EDCs obtained by subtracting the 370 K spectrum, highlighting the disappearance of ZCB at higher temperatures. Spectra are vertically offset for clarity. (c) Temperature dependent MDCs at $E_F$ showing the development of three distinct peaks (marked by green lines) above 350 K, indicative of an electronic structure reorganization. (d) Temperature-dependent MDCs at a binding energy of 135 meV, demonstrating the disappearance of the central peak across the transition. Red lines denote the three bands. (e-f) ARPES intensity maps at 300 K and 370 K, respectively, along the $M$–$\Gamma$–$M$ direction. The color bar represents the ARPES intensity with endpoints marking high (H) and low (L) spectral weigh. (g) EDCs extracted along the yellow and cyan dashed lines shown in (e) and (f), respectively. Vertical black lines in (g) mark the shoulder peaks arising from CDW induced band hybridization. MDCs and EDCs in (c), (d), and \textbf{g} have been normalized to the same height to emphasize temperature-induced changes. EDCs and MDC are integrated within 0.1~\AA$^{-1}$ and 20 meV, respectively. All data were acquired using 92 eV photon energies. 
		}
	\end{figure*}

	\section*{Electronic Reconstruction Across the IC–NC Transition}
	
	To investigate how the structural phase transition influences the low-energy electronic states, we performed temperature-dependent ARPES measurements along the high-symmetry $M$–$\Gamma$–$M$ direction of the Brillouin zone (Fig.~2). At 300 K, well below the NC–IC transition temperature, the electronic band structure displays two highly dispersive branches together with a relatively flat feature centered at $\Gamma$ (hereafter referred to as the zone-centered band, ZCB). Along the dispersive branches, distinct local energy gaps are observed, consistent with the Brillouin-zone folding and band hybridization induced by the periodic lattice distortion in the CDW phase.
	
	Upon warming toward 350~K, the band structure undergoes a progressive evolution. The hybridization gaps weaken and become less defined, reflecting a gradual reduction of the CDW-induced periodic potential. At the same time, the ZCB loses coherence, with its spectral weight redistributed and its peak intensity strongly reduced above the transition. The resulting electronic structure retains only a weak remnant of the $\Gamma$ pocket at the Brillouin-zone center. These changes indicate a suppression of lattice modulation and a reconstruction of the band structure toward a more weakly modulated state resembling the high-temperature phase. As we discuss in the following section, this transformation appears momentum selective, with the most pronounced effects occurring at $\Gamma$, while other regions of the Brillouin zone remain comparatively less affected.

	\section*{Spectral Weight Loss Across the Transition}
	
	To probe the evolution of states near $E_F$, we analyze the energy and momentum distribution curves (EDCs, MDCs) in Fig.~3. EDCs at $\Gamma$ show a marked depletion of ZCB spectral weight across 350 K, signaling coherence loss during the transition (Fig.~3(a)). This is further highlighted in the difference spectra, which clearly reveal the vanishing of the quasiparticle peak in the high-temperature phase (Fig.~3(b)).
	
	The MDCs taken at the the E$_F$ offer a complementary perspective (Fig.~3(c)). At 300 K, they show a broad peak at $\Gamma$, dominated by ZCB weight with contributions from nearby conduction bands. With increasing temperature, this peak narrows and reshapes, indicating redistribution of spectral weight. Above $\sim$350 K, the MDC develops into three peaks—one at $\Gamma$ and two at finite momenta—showing that most coherent ZCB weight has been lost, leaving only a weak central contribution.
	
	To further clarify how this reorganization evolves with energy, we examine MDCs at a deeper binding energy of 135 meV (Fig.~3(d)). The MDCs, here, also display three peaks, but above the transition temperature the central $\Gamma$-point peak disappears. This suggest that ZCB now contributes only marginally at $\Gamma$, with the spectral weight near $E_F$ predominantly carried by the side bands. With ZCB contributing only weakly at $E_F$, conventional scattering-rate analyses becomes impractical; instead, EDCs reveal a progressive loss of coherence, consistent with incoherent carrier redistribution and momentum-selective decoherence.

	Beyond the zone-center states, it is also instructive to examine the dispersive conduction band near $k_{L}$ and $k_{R}$ to assess the Fermi edge behavior across the transition (Figs.~3(e–g)). The corresponding EDCs show no evidence of gap opening: the Fermi edge remains nearly unchanged between 300 K and 370 K. At low temperature, however, distinct shoulder features are observed, arising from CDW-induced band folding and hybridization (Fig.~3(g)). Their disappearance at 370 K signals a strong suppression of CDW order and the associated hybridization gap. While this may be regarded as a weak pseudogap, the finite spectral weight at $E_F$ ensures the system remains metallic at 300 K and above. In this regime, thermal activation overcomes any pseudo-insulating tendency, emphasizing that the resistivity anomaly at $\approx$350 K is governed primarily by coherence–incoherence crossover rather than gap formation.

	\begin{figure*}[t]
		\centering
		\includegraphics[width=0.9\textwidth]{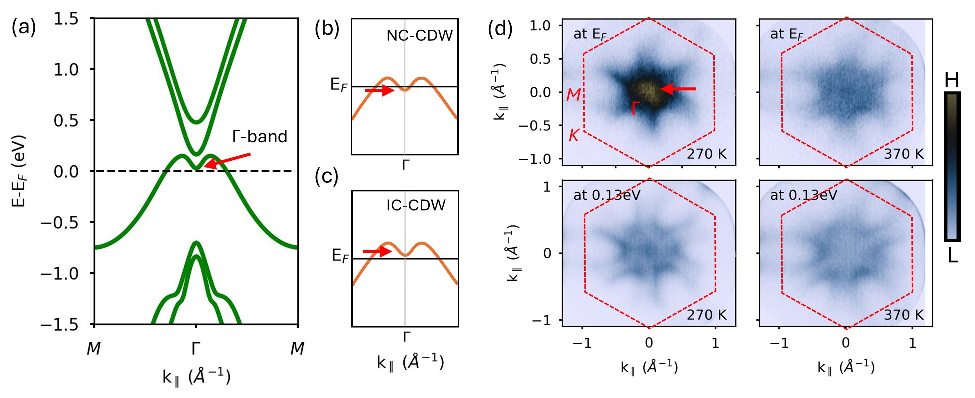}
		\caption{
			(a), Density functional theory band structure of the normal phase. Temperature is accounted for by applying a Fermi–Dirac smearing of $\approx 0.00285$~Ry, corresponding to 450~K. 
			(b), (c), Schematic representation of the conduction band near $\Gamma$ in the NC-CDW and IC-CDW phases, respectively. 
			(d), ARPES constant-energy maps at $E_F$ (top row) and at 0.13~eV binding energy (bottom row), measured at 270~K and 370~K. The data were acquired using 92 eV photon energies.
			At 270~K, the $\Gamma$-centered band (ZCB like feature, marked by arrows) contributes strong spectral weight at $E_F$, whereas at 370~K this weight is strongly reduced and the feature appears only as a weak remnant. This temperature-driven reconstruction reflects the loss of quasiparticle coherence at the NC–IC transition while maintaining continuity with the normal-phase $\Gamma$ conduction state. The color bar represents the ARPES intensity with endpoints marking high (H) and low (L) spectral weigh.}
	\end{figure*}

	\section*{Microscopic mechanism of the NC–IC CDW transition}  
	
	Our ARPES results show that the NC–IC transition is governed by a reorganization of low-energy states at $\Gamma$ without the opening of a full gap. In the NC-CDW phase, the ZCB contributes measurable intensity at $E_F$, whereas in the IC-CDW phase this weight is strongly reduced. Rather than a complete disappearance, this reflects a progressive loss of coherence and redistribution of spectral weight, with metallicity preserved even as the CDW evolves from the NC to the IC state.

	Comparison with the normal-phase band structure provides further context. A ZCB like feature is also found just above $E_F$ in the high-temperature phase (Fig.~4(a)), offering a natural reference point for interpreting the experimental spectra. While CDW reconstruction is expected to modify the details of this state, its continuity across phases and its consistent momentum-space location suggest a common origin. In the NC-CDW phase, hybridization and charge redistribution enhance its contribution at $E_F$, whereas in the IC-CDW phase reduced ordering and weaker screening suppress it, leaving most of the spectral weight above $E_F$ (Figs.~4(b–c)). Correspondingly, ARPES intensity maps (Fig.~4(d)) show pronounced zone-center spectral weight at 270~K evolving into only a weak remnant by 370~K.

	This electronic structure evolution may be linked to nanoscale electronic inhomogeneity. In the NC-CDW state, commensurate clusters coexist with metallic domain walls or discommensurations~\cite{ritschel2015orbital,cho2016nanoscale,ma2016metallic,park2021zoology,salzmann2023observation,geng2023hysteretic}. Preferential charge accumulation in these regions generates local potentials that stabilize the ZCB. Upon entering the IC phase, reduced inhomogeneity and weaker intercluster charge transfer weaken this stabilization, yielding an electronic structure that approaches the high-temperature phase while retaining residual CDW character. These spectroscopic insights suggest that nanoscale phase separations may play an important role in the NC–IC transition in 1T-TaS$_2$, and they provide a microscopic basis for understanding its unconventional transport and nonequilibrium behavior.

	\section*{Correlation between real-space and momentum-space signatures}
	
	To provide a microscopic structural reference for the CDW modulation, Fig.~5(a) illustrates the Star-of-David reconstruction in a single TaS$_2$ layer, where in-plane radial displacements of Ta atoms of order $\sim$0.2,\AA\ form the $\sqrt{13}\times\sqrt{13}$ commensurate superlattice~\cite{rossnagel2011origin}. This distortion underlies the periodic lattice modulation observed in STM and the band folding effects seen in ARPES. To relate the momentum-space behavior to the underlying microscopic texture, we complement our temperature-dependent ARPES measurements with real-space imaging by scanning tunneling microscopy (STM). Although the STM data represent a single-temperature snapshot acquired at 300~K (within the NC-CDW phase), they provide a critical real-space context for interpreting the temperature evolution observed by ARPES.
	
	Figs.~5(b) and 5(c) show STM topographs taken at two different locations on the same cleaved surface. Both regions exhibit a periodic atomic-scale corrugation, but the image contrast varies from place to place, indicating nanoscale variations in the local electronic structure or surface environment. The spatial variations in CDW contrast observed in our STM data are consistent with earlier STM studies of 1T‑TaS$_2$ performed at room temperature~\cite{kim1996temperature}, as well as more recent variable‑temperature STM observations of emergent spatial textures across the CCDW/TCDW/NCCDW phases~\cite{kim1996temperature,geng2023hysteretic}. In addition, low‑temperature STM/STS measurements have revealed that nanoscale electronic inhomogeneities arising from native defects are widespread even in the C‑CDW phase, further supporting the intrinsic, spatially nonuniform electronic landscape of 1T‑TaS$_2$~\cite{campbell2024nanoscale}. Furthermore, Fig.~5(d) displays the height profile extracted along the red line in Fig.~5(c). The oscillations of a few hundred picometers confirm the periodic lattice modulation and further highlight the local contrast variations visible in the topographs. These spatial differences demonstrate that the surface is not perfectly uniform on the nanoscale, forming a natural real-space counterpart to the momentum-space signatures probed by ARPES.
	
	While STM was not performed across the 350~K transition due to experimental constraints (surface stability and thermal drift), the 300~K data reveal intrinsic short-range electronic and structural inhomogeneity already present in the NC phase. As the system approaches the NC--IC transition, the long-range CDW order progressively weakens, and it is expected that fluctuating IC correlations become more dominant. Consequently, the reduction of structural coherence should have its most pronounced impact across this temperature range.
	
	This real-space picture provides a natural framework for interpreting the momentum-space decoherence observed by ARPES near the NC-IC boundary. Across the temperature range where ARPES detects a strong reduction in quasiparticle spectral weight, particularly near the $\Gamma$ point, the increasing role of structural or electronic inhomogeneity likely contributes to the loss of coherence. While the STM measurements are static, they demonstrate that the surface possesses a spatially varying electronic environment. This inherent inhomogeneity suggests that thermal evolution enhances these fluctuations, leading to the observed temperature-driven decoherence in ARPES without necessitating the opening of a conventional insulating gap.

	\begin{figure}[t]
		\centering
		\includegraphics[width=0.45\textwidth]{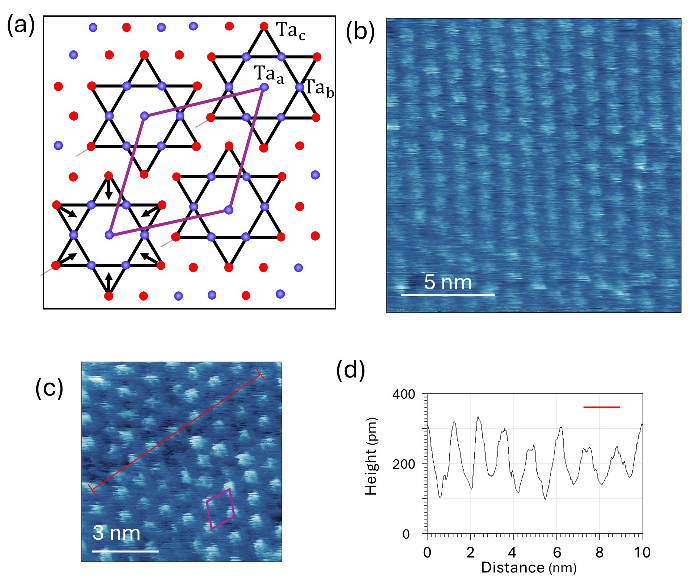}
		\caption{
			(a) Schematic of a single TaS$_2$ layer illustrating the Star-of-David distortion. Only Ta atoms are shown; S atoms are omitted. Arrows indicate in-plane radial displacements from the undistorted lattice (not to scale). The magenta rhombus denotes the commensurate CDW unit cell. The $\sqrt{13}\times\sqrt{13}$ superlattice contains three inequivalent Ta sites, labeled `a', `b', and `c', within the Star-of-David cluster. (b) Large-area STM topography (15 nm $\times$ 15 nm) at Spot~1 showing a well-ordered lattice modulation.  
			(c) High-resolution STM image (9 nm $\times$ 9 nm) at Spot~2 revealing enhanced contrast variations. The magenta rhombus marks a representative atomic arrangement, outlining the $\sqrt{13} \times \sqrt{13}$ CDW supercell with side length $\approx 12$\,\AA. The red line indicates the line profile in (d). STM images emphasize that the observed maxima correspond to the CDW superstructure. (d) Height profile along the red line in (c), showing atomic-scale corrugations of a few hundred picometres. STM data in (b) and (c) were acquired at different surface locations, highlighting spatial variations. Sample bias and tunneling current were 75 mV and 200 pA, respectively. Although the images in (b-c) were obtained at different surface locations, they were selected to illustrate the typical range of local CDW ordering observed in our measurements. The contrast difference therefore reflects representative spatial variations rather than imaging conditions.}
		
	\end{figure}

	\section*{Discussion and Conclusion}  
	
	The $\sim$350 K transition is consistent with contributions from changes in electronic coherence, suggesting that electronic effects may play a role alongside lattice-driven processes. The resistivity anomaly coincides with a clear depletion of quasiparticle spectral weight at $\Gamma$ without gap formation, indicating that the increased resistivity arises from a loss of electronic coherence—likely through enhanced scattering or momentum-selective decoherence.  
	
	These findings connect to STM studies~\cite{cho2016nanoscale,ma2016metallic,stojchevska2014ultrafast}, which reported a nanoscale mosaic of commensurate CDW clusters embedded in metallic-like domain walls. Such spatial inhomogeneity implies that coherence in the NC-CDW phase is intrinsically short-ranged, consistent with the partial decoherence observed in our ARPES spectra. The momentum-space manifestations of decoherence are therefore the spectroscopic counterpart of the real-space fragmentation captured by STM. 
	
	Because these spectral changes are electronically driven, they should be highly susceptible to ultrafast control. Indeed, femtosecond optical or electrical pulses have already been shown to induce hidden states and nonthermal switching in 1\textit{T}-TaS$_2$~\cite{vaskivskyi2015controlling,stojchevska2014ultrafast}. The absence of an insulating gap means that carrier delocalization can be triggered with minimal energy cost, an advantage for low-power, high-speed switching. This may have implications for phase-change and resistivity-switching concepts operating near ambient conditions.  
	
	In summary, our study provides a microscopic view of phase evolution in 1\textit{T}-TaS$_2$, combining momentum-resolved spectroscopy with insights into nanoscale electronic texture. By linking real-space inhomogeneity to momentum-space decoherence, our results contribute to the understanding of CDW physics and highlight opportunities for controlling electronic phases in layered quantum materials, with particular promise for ultrafast switching and resistivity-memory devices.

	\vspace{0.5cm}
	\noindent
	\textbf{METHODS} \\

	\textbf{Materials}
	\noindent
	
	The 1T-TaS$_2$ samples used in this study were commercially obtained bulk single crystals from 2dsemiconductors, with typical lateral dimensions between 0.5~cm and 1~cm. For ARPES and STM experiments, the crystals were mounted on Al sample plates using Ag epoxy, and a ceramic post was affixed to the top surface with Ag epoxy. Cleaving was performed \textit{in situ} under ultra-high vacuum conditions by knocking off the ceramic post with a wobble stick inside the ARPES or STM chamber. To ensure reliable surface preparation, we typically used smaller pieces of $\sim$1~mm $\times$ 1~mm crystals for cleaving. This procedure consistently produced fresh, atomically clean surfaces without the need for additional chemical treatment~\cite{zhang2022angle}.\\
	
	\noindent
	\textbf{Micro angle resolved photoemission spectroscopy}
	
	\noindent
	Micro-ARPES measurements were carried out at the 21-ID-1 (ESM) beamline of the National Synchrotron Light Source II, Brookhaven National Laboratory~\cite{rajapitamahuni2024electron}. The setup employs a Scienta DA30 analyzer equipped with deflector mode, which allows full Fermi surface mapping without mechanically rotating the sample. High-quality 1\textit{T}-TaS$_2$ single crystals were fixed to copper holders using silver epoxy and cleaved \textit{in situ} under ultra-high-vacuum conditions with a base pressure below $5 \times 10^{-11}$~Torr. And the incident photon beam was directed onto the sample at an angle of $55^\circ$ with respect to the surface.\\
	
	\noindent
	\textbf{Transport measurements}
	Temperature dependent resistance measurements were performed using the dc resistivity option in Evercool Physical Property Measurement System (Quantum Design Inc.). The data was recorded in sweep mode while both heating and cooling the sample at the rate of 1 K/min.\\
	
	\noindent
	\textbf{Low energy electron diffraction experiment}
	LEED/$\mu$-LEED measurements were performed at the PEEM/LEEM end-station (beamline BL3.2Ub) of the Synchrotron Light Research Institute (SLRI), Nakhon Ratchasima, using an Elmitec SPELEEM-III microscope. Samples were cleaved under ultrahigh vacuum (UHV, $p \approx 10^{-10}$-$10^{-9}$\,mbar). Diffraction patterns were recorded under near-normal incidence with the instrument's MCP/CCD detector; for $\mu$-LEED, patterns were acquired from $\sim$1-2\,$\mu$m regions.\\

	\noindent
	\textbf{Scanning Tunneling Microscopy}
	\noindent
	STM measurements were performed using an Omicron VT-STM XA~650 system operated under ultrahigh vacuum (UHV) conditions with a base pressure of $2 \times 10^{-10}$~Torr at room temperature. The samples cleaved in the UHV system. All STM images were acquired in constant-current mode using Pt/Ir tips, and all bias voltages mentioned in the text correspond to the bias applied to the sample. The STM data were processed and analyzed using the Gwyddion software package.\\

	\noindent
	\textbf{Density functional theory}
	
	\noindent
	Electronic structure calculations were carried out within the plane-wave density functional theory framework as implemented in the \textsc{Quantum ESPRESSO} package~\cite{giannozzi2009quantum}. The exchange–correlation functional was described using the generalized gradient approximation of Perdew, Burke, and Ernzerhof (PBE)~\cite{perdew1996generalized}. For the ionic potentials, fully relativistic norm-conserving pseudopotentials from the PseudoDojo library were employed~\cite{van2018pseudodojo,hamann2013optimized}. Plane-wave basis sets were truncated at kinetic energy cutoffs of 80~Ry for the wavefunctions and 500~Ry for the charge density. Sampling of the Brillouin zone for bulk 1\textit{T}-TaS$_2$ in the normal phase utilized an $11 \times 11 \times 11$ Monkhorst–Pack $k$-point grid.\\
	
	\noindent
	\textbf{Data availability}
	
	\noindent
	The data that support the findings of this study are available from the corresponding
	author upon request. \\
	
	\noindent
	\textbf{Acknowledgments}
	
	\noindent
	This research used resources ESM (21ID-I) beamline of the National Synchrotron Light Source II, a U.S. Department of Energy (DOE) Office of Science User Facility operated for the DOE Office of Science by Brookhaven National Laboratory under Contract No. DE-SC0012704. This work also utilized the resources of the Center for Functional Nanomaterials at Brookhaven National Laboratory, supported by the U.S. Department of Energy, Office of Basic Energy Sciences, under Contract No.~DE-SC0012704. We have no conflict of interest, financial or other to declare. Author MJ would like to acknowledge the support from NSF grant (award number 2233149) and from University of Connecticut Quantum Seed grant. Author J.C. notes that this research also benefited from support provided by the Xiamen University Malaysia research grant (Grant No.~IORI/0007).

	\noindent
	\textbf{Author Contributions}
	\noindent
	
	T.Y. conceived and supervised the research. T.Y., A.R., A.K.K., and E.V. conducted the photoemission experiments at the 21-ID-1 beamline. X.T. performed the STM experiments. T.W. and P.T. conducted LEED data. M.J. carried out the transport measurements. Y.S.N. performed the theoretical calculations under the supervision from J.C.Z. T.Y. wrote the manuscript with input from all authors. All authors contributed to the scientific discussion and interpretation of the results.
	
	\noindent
	\textbf{Competing Interests}
	\noindent
	
	The authors declare no competing interests.

\end{document}